\documentclass{article}
\usepackage{amssymb}
\usepackage{amsmath}

\input{tcilatex}

\begin{document}

\title{\textbf{Finslerian Quantum Field Theory}}
\author{Howard E. Brandt \\
U.S. Army Research Laboratory, Adelphi, Maryland, U.S.\\
hbrandt@arl.army.mil}
\maketitle

\begin{abstract}
Finsler geometry motivates a generalization of the Riemannian structure of
spacetime to include dependence of the spacetime metric and associated
invariant tensor fields on the four-velocity coordinates as well as the
spacetime coordinates of the observer. It is then useful to consider the
tangent bundle of spacetime with spacetime in the base manifold and
four-velocity space in the fiber. A physical basis for the differential
geometric structure of the spacetime tangent bundle is provided by the
universal upper limit on proper acceleration relative to the vacuum. It is
then natural to consider a quantum field having a vanishing eigenvalue when
acted on by the Laplace-Beltrami operator of the spacetime tangent bundle.
On this basis a quantum field theory can be constructed having a built-in
intrinsic regularization at the Planck scale, and finite vacuum energy
density.
\end{abstract}

\section{\protect\bigskip FINSLERIAN FIELDS}

A physical Finslerian field $F(x.v)$ is one that depends not only on the
observer's spacetime coordinates, 
\begin{equation}
x\equiv \{x^{\mu }\}=\{x^{0},x^{1},x^{2},x^{3}\},
\end{equation}%
but also on the observer's four-velocity coordinates,%
\begin{equation}
v\equiv \{v^{\mu }\}=\{dx^{\mu }/ds\}=\{v^{0},v^{1},v^{2},v^{3}\},
\end{equation}%
where $ds$ is the infinitesimal interval along the worldline of the observer 
\cite{Brandt1}-\cite{Brandt3}. The four-velocity coordinates play here the
role of the tangent space coordinates of Finsler geometry. It can be argued
that the spacetime-metric field $g_{\mu \nu }$ itself must in general depend
not only on where it is observed in spacetime, but also on the four-velocity
of the `observer', namely it is a Finslerian field \cite{Brandt1}-\cite%
{Brandt3}: 
\begin{equation}
g_{\mu \nu }=g_{\mu \nu }(x,v).
\end{equation}%
(The reader may prefer to replace the word `observer' by `measuring device',
`object acted upon by the field', or `some other field interacting locally
with the field'.) The spacetime metric in a canonical Finsler spacetime is
not only Finslerian, but also satisfies special homogeneity conditions
involving the dependence of the metric on the tangent space coordinates, $v.$
In particular, one has \cite{Brandt4}, \cite{Rund} 
\begin{equation}
ds=L(x,dx),
\end{equation}%
where $L$ is the fundamental Finsler function, and 
\begin{equation}
L(x,adx)=aL(x,dx),
\end{equation}%
from which it follows that 
\begin{equation}
g_{\mu \nu }=\frac{1}{2}\frac{\partial ^{2}}{\partial v^{\mu }\partial
v^{\nu }}L^{2}(x,v),
\end{equation}%
\begin{equation}
L^{2}(x,v)=g_{\mu \nu }v^{\mu }v^{\nu },
\end{equation}%
\begin{equation}
v^{\alpha }\frac{\partial }{\partial v^{\alpha }}g_{\mu \nu }=v^{\alpha }%
\frac{\partial }{\partial v^{\mu }}g_{\alpha \nu }=0,
\end{equation}%
and 
\begin{equation}
\frac{\partial }{\partial v^{\alpha }}g_{\mu \nu }=\frac{\partial }{\partial
v^{\mu }}g_{\alpha \nu }.
\end{equation}%
In considering \ Finslerian spacetime and associated embedded Finslerian
fields, it is useful to consider the tangent bundle of spacetime with
spacetime in the base manifold and four-velocity space in the fiber. Using
the homogeneity relations, Eqs.(4)-(9), then the connection and Riemann
curvature scalar of the spacetime tangent bundle \cite{Brandt5} can be
significantly reduced for the case of a Finsler-spacetime base manifold \cite%
{Brandt4}, \cite{Yano}. However, the special homogeneity requirements may
not hold physically in general, but may only hold in certain special
spacetime models.

A physical basis for the differential geometric structure of the spacetime
tangent bundle is provided by the universal upper limit $a_{0}$ on proper
acceleration $a$ relative to the vacuum \cite{Brandt6}-\cite{Brandt9}. If
the proper acceleration were sufficiently large, then, because of vacuum
radiation in an accelerated frame (in which particles are produced with
average energy proportional to the proper acceleration), particles would be
produced with mass such that their Schwarzschild radius exceeds their extent
(Compton wavelength). Copious production of black-hole anti-black-hole pairs
would ensue, accompanied by breakdown of the classical spacetime structure,
and the very concept of acceleration would loose any meaning because of the
resulting complex topology. Explicitly, the maximal proper acceleration, $%
a_{0}$, is given by \cite{Brandt6} 
\begin{equation}
a_{0}=2\pi \alpha \left( \frac{c^{7}}{\hslash G}\right) ^{1/2},
\end{equation}%
where $\alpha $ is a number of order unity, $c$ is the speed of light, $%
\hslash $ is Planck's constant divided by 2$\pi $, and $G$ is the universal
gravitational constant. This is the maximum possible proper acceleration
relative to the vacuum and is taken to be universal. Hence for any proper
acceleration $a$, one requires%
\begin{equation}
a^{2}\leq a_{0}^{2}.
\end{equation}%
But, according to the differential geometry of spacetime, the proper
acceleration, $a$, along a worldline in curved spacetime is given by 
\begin{equation}
a^{2}=-c^{4}g_{\mu \nu }\frac{Dv^{\mu }}{ds}\frac{Dv^{\nu }}{ds},
\end{equation}%
where the four-velocity $v^{\mu }$ is given by 
\begin{equation}
v^{\mu }=\frac{dx^{\mu }}{ds},
\end{equation}%
and $\frac{Dv^{\mu }}{ds}$ denotes the covariant derivative of the
four-velocity with respect to the interval along the worldline, namely, 
\begin{equation}
\frac{Dv^{\mu }}{ds}=\frac{dv^{\mu }}{ds}+\Gamma _{\ \alpha \beta }^{\mu
}v^{\alpha }v^{\beta },
\end{equation}%
in which $\Gamma _{\ \alpha \beta }^{\mu }$ is the spacetime affine
connection, and 
\begin{equation}
ds^{2}\equiv g_{\mu \nu }dx^{\mu }dx^{\nu }
\end{equation}%
is the line element of spacetime. Then substituting Eqs.(12) and (14) in
Eq.(11), one obtains 
\begin{equation}
-c^{4}g_{\mu \nu }\left( \frac{dv^{\mu }}{ds}+\Gamma _{\ \alpha \beta }^{\mu
}v^{\alpha }v^{\beta }\right) \left( \frac{dv^{\nu }}{ds}+\Gamma _{\ \lambda
\delta }^{\nu }v^{\lambda }v^{\delta }\right) \leq a_{0}^{2}.
\end{equation}%
Next substituting Eqs.(13) and (15) in Eq.(16), one obtains \cite{Brandt8} 
\begin{equation}
d\sigma ^{2}\equiv g_{\mu v}dx^{\mu }dx^{\nu }+\rho _{0}^{2}g_{\mu v}\left(
dv^{\mu }+\Gamma _{\ \alpha \beta }^{\mu }v^{\alpha }dx^{\beta }\right)
\left( dv^{\nu }+\Gamma _{\ \lambda \delta }^{\nu }v^{\lambda }dx^{\delta
}\right) \geq 0,
\end{equation}%
where 
\begin{equation}
\rho _{0}=\frac{c^{2}}{a_{0}},
\end{equation}%
is the minimum radius of curvature of worldlines. Equation (17) defines the
eight-dimensional quadratic form $d\sigma ^{2}$, which is nonnegative along
the worldline. The inequality, Eq. (17), simply expresses the fact that the
proper acceleration can never exceed the maximal proper acceleration. By
analogy with the construction of the spacetime line element of general
relativity from the limiting speed of light, it is natural to take $d\sigma
^{2}$ to be the line element in the tangent bundle of spacetime, in which
the spacetime coordinates $x^{\mu }$ are the coordinates in the spacetime
base manifold, and the four-velocity coordinates $\rho _{0}v^{\mu }$ (modulo
a factor of $\rho _{0}$) are the tangent space coordinates.

The bundle line element, Eq.(17), can be rewritten as follows \cite{Brandt8}%
, \cite{Brandt1}:%
\begin{equation}
d\sigma ^{2}\equiv G_{MN}dx^{M}dx^{N},\ \ \ \ \{M,N=0,2,...,7\},
\end{equation}%
where the bundle coordinates are 
\begin{equation}
\left\{ x^{M}\right\} \equiv \left\{ x^{\mu },\rho _{0}v^{\mu }\right\} ,\ \
\ \{M=0,2,...,7;\ \mu =0,1,2,3\},
\end{equation}%
and the metric of the tangent bundle of spacetime is 
\begin{equation}
G_{MN}=\left[ 
\begin{array}{cc}
g_{\mu \nu }+g_{\alpha \beta }A_{\ \mu }^{\alpha }A_{\ \nu }^{\beta } & 
A_{n\mu } \\ 
A_{m\nu } & g_{mn}%
\end{array}%
\right] ,
\end{equation}%
in which%
\begin{equation}
A_{\ \nu }^{\mu }=\rho _{0}v^{\lambda }\Gamma _{\ \lambda \nu .}^{\mu }
\end{equation}%
The bundle metric $G_{MN}$, given by Eq.(21), has a structure similar to
that of an eight-dimensional Kaluza-Klein gauge theory in which the higher
dimensions are in four-velocity space, and $A_{\ \nu }^{\mu }$ is the gauge
potential. Eqs.( 19)-(22) served as the starting point for investigating
possible implications of a limiting proper acceleration for the differential
geometric structure of the tangent bundle of spacetime \cite{Brandt1}-\cite%
{Brandt4}, \cite{Brandt5}, \cite{Brandt6}-\cite{Brandt22}. Possible forms
for the bundle connection, curvature, and action were explored, including
those based on Riemannian and Finsler spacetimes, and also K\"{a}hler and
complex spacetime tangent bundles. Among the many differential geometric
invariants of the spacetime tangent bundle, important for the present
discussions is the Laplace-Beltrami operator:%
\begin{equation}
\mathcal{L}=G^{-1/2}\frac{\partial }{\partial x^{M}}\left( G^{1/2}G^{MN}%
\frac{\partial }{\partial x^{N}}\right) .
\end{equation}%
This is the invariant generalization of the wave operator, or d'Alembertian,
of field theory. A simple invariant field equation for a Finslerian scalar
field $\phi (x,v)$ is then given by \cite{Brandt18} 
\begin{equation}
\mathcal{L\phi (}x,v)=0.
\end{equation}

\section{FINSLERIAN SCALAR QUANTUM FIELD}

When the spacetime is Minkowskian, the ordinary inhomogeneous Lorentz group
(or Poincar\'{e} group) is a subgroup of the invariance group of the
spacetime tangent bundle \cite{Brandt10}, \cite{Brandt18}. It is of interest
to examine quantum field solutions to Eq.(24) for this simple case, in order
to establish connections between the Finslerian framework and canonical
relativistic quantum field theory. To this end, consider a simple case in
which the spacetime is completely flat. In particular, take the spacetime
metric to be Minkowskian. For this case, it was argued in earlier work that
the scalar field satisfying Eq.(24) is given by \cite{Brandt18}, \cite%
{Brandt23}, \cite{Brandt24}:%
\begin{equation}
\begin{array}{c}
\phi (x,v)=\int \frac{d^{3}\mathbf{p}}{\left( 2\pi \hbar \right)
^{3/2}\left( 2p^{0}N\right) ^{1/2}}\left[ e^{-ipx/\hbar }e^{-\rho
_{0}pv/\hbar }\theta _{1}(\rho _{0}pv/\hbar )a(\mathbf{p})\right. \\ 
+\;\left. e^{ipx/\hbar }e^{\rho _{0}pv/\hbar }\theta _{1}(-\rho _{0}pv/\hbar
)a^{\dagger }(\mathbf{p})\right] ,%
\end{array}%
\end{equation}%
where $p$ denotes the four-momentum $p^{\mu }=\{p^{0},p^{1},p^{2},p^{3}\}$
of a particle excitation of the scalar field, $a^{\dagger }(\mathbf{p})$ and 
$a(\mathbf{p})$ are the particle creation and annihilation operators
satisfying the commutation relations%
\begin{equation}
\lbrack a(\mathbf{p}),a^{\dagger }(\mathbf{p}^{\prime })]=\delta ^{3}(%
\mathbf{p}-\mathbf{p}^{\prime }),\ \text{ }[a(\mathbf{p}),a(\mathbf{p}%
^{\prime })]=0,\ \text{\ }[a^{\dagger }(\mathbf{p}),a^{\dagger }(\mathbf{p}%
^{\prime })]=0,
\end{equation}%
$\delta ^{3}(\mathbf{z})$ is the three-dimensional Dirac delta function, and 
$\theta _{1}\left( z\right) $ is the Heaviside function \cite{Kanwal},%
\begin{equation}
\theta _{1}\left( z\right) =\left\{ 
\begin{array}{c}
1,\;\;z\geq 0 \\ 
0,\;\;z<0%
\end{array}%
\right. .
\end{equation}%
Also in Eq.(25), $N$ is a normalization factor such that the field operator
acting on the vacuum state, namely $\left\langle 0\right\vert \phi (x,v)$,
is the state of the field corresponding to a single-particle excitation
appearing at some point $(x,v)$ in the spacetime tangent bundle, namely \cite%
{Brandt23}, \cite{Brandt24},%
\begin{equation}
N=\frac{\rho _{0}V}{16\pi ^{2}\hbar ^{2}m^{2}c^{2}}K_{1}\left( 2\rho
_{0}mc/\hbar \right) ,
\end{equation}%
where $V$ denotes the volume of space, $m$ is the mass of the scalar
particle excitation of the quantum field, and $K_{1}(z)$ is the modified
Bessel function of the third kind of order one.

It can be shown that both the positive and negative frequency terms in
Eq.(25) are proportional to \cite{Brandt18}, \cite{Brandt7}:%
\begin{equation}
e^{-\rho _{0}\left\vert pv\right\vert /\hbar }=\exp \left\{ -\frac{1}{2\pi
\alpha }\left( \frac{G}{\hbar c}\right) ^{1/2}\gamma m\left( \left[ 1+\left( 
\frac{\left\vert \mathbf{p}\right\vert }{mc}\right) ^{2}\right] ^{1/2}-\frac{%
\mathbf{p\cdot }d\mathbf{x}/dt}{mc^{2}}\right) \right\} ,
\end{equation}%
where $d\mathbf{x}/dt$ is the ordinary velocity of the observer, and 
\begin{equation}
\gamma =\left[ 1-\left( \frac{\left\vert d\mathbf{x}/dt\right\vert }{c}%
\right) ^{2}\right] ^{-1/2}.
\end{equation}%
As can be seen from Eq.(29), there occurs an intrinsic Planck-scale
regularization of the quantum field, with an exponential energy cutoff
beyond the Planck energy.

It is important to note that for an observer with ordinary velocity 
\begin{equation}
\overset{\cdot }{\mathbf{x}}\ \equiv d\mathbf{x}/dt,
\end{equation}%
in Minkowski spacetime, one has 
\begin{equation}
ds=\pm \sqrt{\left( dx^{0}\right) ^{2}-\left\vert d\mathbf{x}\right\vert ^{2}%
}=\pm \frac{dx^{0}}{\gamma },
\end{equation}%
and therefore 
\begin{equation}
v^{\mu }\equiv \frac{dx^{\mu }}{ds}=\pm \{\gamma ,\gamma \overset{\cdot }{%
\mathbf{x}}/c\}.
\end{equation}%
Thus the ordinary velocity $\overset{\cdot }{\mathbf{x}}$ corresponds to
both future and past directed four velocity. This is essential in order that
both the positive-frequency particle modes and the negative-frequency
antiparticle modes in Eq.(25) and in the following have nonvanishing
support. One is reminded of the St\"{u}ckelberg-Feynman idea that
antiparticles can be represented as particles with proper time reversed
relative to true time \cite{Brandt19}, \cite{Feynman}.

\section{HAMILTONIAN OF SCALAR QUANTUM FIELD}

On the basis of microcausality, it was argued in earlier work \cite{Brandt19}
that it is logical to define the generalized adjoint $\left( \phi
(x,v)\right) ^{\dagger }$ of the quantum field $\phi (x,v)$ by 
\begin{equation}
\left( \phi (x,v)\right) ^{\dagger }=\phi ^{\dagger }(x,-v),
\end{equation}%
in which the ordinary adjoint is taken, but also the sign of the
four-velocity $v$ is changed. This is connected, through charge-conjugation,
with the fact that the positive-frequency particle component of the field in
Eq.(25) has nonvanishing support for positive $v^{0}$ (or equivalently,
positive $pv$), and the negative- frequency anti-particle component has
nonvanishing support for negative $v^{0}$ (equivalently, negative $pv$) \cite%
{Brandt19}. In order that the Hamiltonian be Hermitian (in terms of the
generalized adjoint), that it reduce to the canonical form in Fock space,
and that the observer's four-velocity lie on the four-velocity shell ($%
v^{2}=1)$, it is natural to define the Hamiltonian for the scalar quantum
field as follows \cite{Brandt23}, \cite{Brandt24}:%
\begin{equation}
H=\int \rho _{0}^{4}d^{4}v\delta \mathbf{(}\rho _{0}^{2}v^{2}-\rho
_{0}^{2})d^{3}\mathbf{x}T_{00}(\mathbf{x,}v),
\end{equation}%
in which the bundle energy density $T_{00}(\mathbf{x,}v)$ at point $(x,v)$
in the bundle is integrated over the entire spacetime tangent bundle with
large spatial volume $V$, $\mathbf{\delta (}z)$ is the one-dimensional Dirac
delta function, and the bundle energy density is given by%
\begin{equation}
\begin{array}{c}
T_{00}(\mathbf{x,}v)=\frac{V}{32\pi ^{3}\hslash m}\left[ \frac{\partial }{%
\partial x^{0}}\phi \left( \mathbf{x,-}v\right) \frac{\partial }{\partial
x^{0}}\phi \left( \mathbf{x,}v\right) +\nabla _{\mathbf{x}}\phi \left( 
\mathbf{x,-}v\right) \cdot \nabla _{\mathbf{x}}\phi \left( \mathbf{x,}%
v\right) \right.  \\ 
+\left. \frac{1}{\rho _{0}^{2}}\frac{\partial }{\partial v^{\mu }}\phi
\left( \mathbf{x,-}v\right) \frac{\partial }{\partial v_{\mu }}\phi \left( 
\mathbf{x,}v\right) \right] .%
\end{array}%
\end{equation}

To see that the Hamiltonian is Hermitian, one first notes that 
\begin{equation}
H^{\dagger }=\int \rho _{0}^{4}d^{4}v\delta \mathbf{(}\rho
_{0}^{2}v^{2}-\rho _{0}^{2})d^{3}\mathbf{x}T_{00}(\mathbf{x,}v)^{\dagger }.
\end{equation}%
Next, taking the generalized adjoint of $T_{00}(\mathbf{x,}v)$ and using
Eq.(34), one obtains%
\begin{equation}
\begin{array}{c}
T_{00}(\mathbf{x,}v)^{\dagger }=\frac{V}{32\pi ^{3}\hslash m}\left[ \frac{%
\partial }{\partial x^{0}}\phi ^{\dagger }\left( \mathbf{x,}-v\right) \frac{%
\partial }{\partial x^{0}}\phi ^{\dagger }\left( \mathbf{x,}v\right) +\nabla
_{\mathbf{x}}\phi ^{\dagger }\left( \mathbf{x,}-v\right) \cdot \nabla _{%
\mathbf{x}}\phi ^{\dagger }\left( \mathbf{x,}v\right) \right.  \\ 
+\left. \frac{1}{\rho _{0}^{2}}\frac{\partial }{\partial v^{\mu }}\phi
^{\dagger }\left( \mathbf{x,}-v\right) \frac{\partial }{\partial v_{\mu }}%
\phi ^{\dagger }\left( \mathbf{x,}v\right) \right] .%
\end{array}%
\end{equation}%
Equivalently, using Eq.(25), Eq.(38) becomes\qquad 
\begin{equation}
\begin{array}{c}
T_{00}(\mathbf{x,}v)^{\dagger }=\frac{V}{32\pi ^{3}\hslash m}\left[ \frac{%
\partial }{\partial x^{0}}\phi \left( \mathbf{x,}v\right) \frac{\partial }{%
\partial x^{0}}\phi \left( \mathbf{x,}-v\right) +\nabla _{\mathbf{x}}\phi
\left( \mathbf{x,}v\right) \cdot \nabla _{\mathbf{x}}\phi \left( \mathbf{x,}%
-v\right) \right.  \\ 
+\left. \frac{1}{\rho _{0}^{2}}\frac{\partial }{\partial v^{\mu }}\phi
\left( \mathbf{x,}v\right) \frac{\partial }{\partial v_{\mu }}\phi \left( 
\mathbf{x,}-v\right) \right] ,%
\end{array}%
\end{equation}%
or comparing Eqs.(39) and (36), one obtains%
\begin{equation}
T_{00}(\mathbf{x,}v)^{\dagger }=T_{00}(\mathbf{x,}-v).
\end{equation}%
Next substituting Eq.(40) in Eq.(37), one has%
\begin{equation}
H^{\dagger }=\int \rho _{0}^{4}d^{4}v\delta \mathbf{(}\rho
_{0}^{2}v^{2}-\rho _{0}^{2})d^{3}\mathbf{x}T_{00}(\mathbf{x,}-v),
\end{equation}%
and replacing the dummy variable of integration $v$ by $-v$, Eq.(41) becomes%
\begin{equation}
H^{\dagger }=\int \rho _{0}^{4}d^{4}v\delta \mathbf{(}\rho
_{0}^{2}v^{2}-\rho _{0}^{2})d^{3}\mathbf{x}T_{00}(\mathbf{x,}v).
\end{equation}%
Finally, substituting Eq.(35) in Eq.(42), one concludes that 
\begin{equation}
H^{\dagger }=H.
\end{equation}%
Thus $H^{\dagger }$ is in fact Hermitian.

Next, using Eqs.(25), (27) and (36) in Eq.(35), integrating over space, and
substituting the integral \cite{Brandt23}, \cite{Ober}%
\begin{equation}
\int \frac{d^{3}\mathbf{v}}{2\,v^{0}}e^{-2\rho _{0}pv/\hbar }=\frac{\pi
\hbar }{\rho _{0}mc}K_{1}\left( 2\rho _{0}mc/\hbar \right) ,
\end{equation}%
one obtains the canonical expression for the Hamiltonian operator for a
scalar quantum field in Fock space \cite{Brandt23}:%
\begin{equation}
H=\frac{1}{2}c\int d^{3}\mathbf{p}p^{0}\left[ a^{\dagger }(\mathbf{p})a(%
\mathbf{p})+a(\mathbf{p})a^{\dagger }(\mathbf{p})\right] ,
\end{equation}%
in which%
\begin{equation}
p^{0}=\left( m^{2}c^{2}+\left\vert \mathbf{p}\right\vert ^{2}\right) ^{1/2}.
\end{equation}

For consistency, it is also well to verify that the Heisenberg equation of
motion for the field $\phi $ is satisfied, namely \cite{Zee}, 
\begin{equation}
\lbrack H,\phi ]=-i\hbar \frac{\partial }{\partial t}\phi .
\end{equation}%
To see that Eq.(47) holds, one uses Eqs.(45) and (25) to evaluate the
commutator $[H,\phi ]$. Thus one has 
\begin{eqnarray}
\lbrack H,\phi ] &=&\frac{1}{2}c\int d^{3}\mathbf{p}^{\prime }p^{0\prime
}[\left( a^{\dagger }(\mathbf{p}^{\prime })a(\mathbf{p}^{\prime })+a(\mathbf{%
p}^{\prime })a^{\dagger }(\mathbf{p}^{\prime })\right) ,\int \frac{d^{3}%
\mathbf{p}}{\left( 2\pi \hbar \right) ^{3/2}\left( 2p^{0}N\right) ^{1/2}} 
\notag \\
&&\left\{ e^{-ipx/\hbar }e^{-\rho _{0}pv/\hbar }\theta _{1}(\rho
_{0}pv/\hbar )a(\mathbf{p})\right.  \notag \\
&&+\left. e^{ipx/\hbar }e^{\rho _{0}pv/\hbar }\theta _{1}(-\rho _{0}pv/\hbar
)a^{\dagger }(\mathbf{p})\right\} ],
\end{eqnarray}%
or equivalently, 
\begin{eqnarray}
\lbrack H,\phi ] &=&\frac{1}{2}c\int \frac{d^{3}\mathbf{p}^{\prime
}p^{0\prime }d^{3}\mathbf{p}}{\left( 2\pi \hbar \right) ^{3/2}\left(
2p^{0}N\right) ^{1/2}}\left\{ e^{-ipx/\hbar }e^{-\rho _{0}pv/\hbar }\theta
_{1}(\rho _{0}pv/\hbar )\right.  \notag \\
&&\left( [a^{\dagger }(\mathbf{p}^{\prime })a(\mathbf{p}^{\prime }),a(%
\mathbf{p})]+[a(\mathbf{p}^{\prime })a^{\dagger }(\mathbf{p}^{\prime }),a(%
\mathbf{p})]\right)  \notag \\
&&+e^{ipx/\hbar }e^{\rho _{0}pv/\hbar }\theta _{1}(-\rho _{0}pv/\hbar ) 
\notag \\
&&\left. \left( [a^{\dagger }(\mathbf{p}^{\prime })a(\mathbf{p}^{\prime
}),a^{\dagger }(\mathbf{p})]+[a(\mathbf{p}^{\prime })a^{\dagger }(\mathbf{p}%
^{\prime }),a^{\dagger }(\mathbf{p})]\right) \right\} .
\end{eqnarray}%
Then substituting Eqs.(26) in Eq.(49), one obtains%
\begin{eqnarray}
\lbrack H,\phi ] &=&-c\int \frac{d^{3}\mathbf{p}p^{0}}{\left( 2\pi \hbar
\right) ^{3/2}\left( 2p^{0}N\right) ^{1/2}}\left\{ e^{-ipx/\hbar }e^{-\rho
_{0}pv/\hbar }\theta _{1}(\rho _{0}pv/\hbar )a(\mathbf{p})\right.  \notag \\
&&\left. -e^{ipx/\hbar }e^{\rho _{0}pv/\hbar }\theta _{1}(-\rho _{0}pv/\hbar
)a^{\dagger }(\mathbf{p})\right\} .
\end{eqnarray}%
Finally, substituting Eq.(25) in Eq.(50), one obtains the Heisenberg
equation of motion for the field, Eq.(47).

\section{VACUUM ENERGY DENSITY}

Using Eqs.(35) and (36), and denoting the vacuum state by $\left\vert
0\right\rangle $, it follows that the vacuum energy in the bundle, for
spatial volume $V$, is given by \cite{Brandt23}:%
\begin{equation}
\left\langle 0\right\vert H\left\vert 0\right\rangle =\frac{1}{2}cV\int 
\frac{d^{3}\mathbf{p}}{\left( 2\pi \hbar \right) ^{3}}\left(
m^{2}c^{2}+\left\vert \mathbf{p}\right\vert ^{2}\right) ^{1/2},
\end{equation}%
which is the canonical divergent result \cite{Zee}. However, a particular
observer has four-velocity $v$, and his worldline at any time is confined to
the neighborhood of $v$ in the fiber (See Eq.(33)). The vacuum energy which
he observes is therefore given by \cite{Brandt23}%
\begin{equation}
\left\langle 0\right\vert \Delta H\left\vert 0\right\rangle =\rho
_{0}^{3}\Delta ^{3}\mathbf{v}\left\langle 0\right\vert \int \rho
_{0}dv^{0}\delta \mathbf{(}\rho _{0}^{2}v^{2}-\rho _{0}^{2})d^{3}\mathbf{x}%
T_{00}(\mathbf{x,}v)\left\vert 0\right\rangle ,
\end{equation}%
in which $\Delta ^{3}\mathbf{v}$ is defined by 
\begin{equation}
\Delta ^{3}\mathbf{v\equiv \ }v^{0}\int_{\mathbf{v}-\delta \mathbf{v}/2}^{%
\mathbf{v}+\delta \mathbf{v}/2}\frac{d^{3}\mathbf{v}^{\prime }}{v^{0\prime }}%
\underset{\delta \mathbf{v\ll v}}{\longrightarrow }\delta ^{3}\mathbf{v,}
\end{equation}%
where $\delta \mathbf{v}$ is the spread in the spatial components of the
four-velocity of the observer, and because of the Dirac delta function in
Eq.(52),%
\begin{equation}
v^{0\prime }=\left( 1+\left\vert \mathbf{v}^{\prime }\right\vert ^{2}\right)
^{1/2},\ \ v^{0}=\left( 1+\left\vert \mathbf{v}\right\vert ^{2}\right)
^{1/2}.
\end{equation}%
One notes that in Eq.(52), one can use the well known identity:%
\begin{eqnarray}
&&\delta \mathbf{(}\rho _{0}^{2}v^{2}-\rho _{0}^{2})  \notag \\
&=&\frac{1}{\rho _{0}^{2}}\delta \mathbf{(}\{v^{0}+\left[ 1+\left\vert 
\mathbf{v}\right\vert ^{2}\right] ^{1/2}\}\{v^{0}-\left[ 1+\left\vert 
\mathbf{v}\right\vert ^{2}\right] ^{1/2}\}) \\
&=&\frac{1}{2\rho _{0}^{2}\left[ 1+\left\vert \mathbf{v}\right\vert ^{2}%
\right] ^{1/2}}\left[ \delta \left( v^{0}-\left[ 1+\left\vert \mathbf{v}%
\right\vert ^{2}\right] ^{1/2}\right) +\delta \left( v^{0}+\left[
1+\left\vert \mathbf{v}\right\vert ^{2}\right] ^{1/2}\right) \right] . 
\notag
\end{eqnarray}%
Equation (53), defining the spread, is true because the expression following 
$\Delta ^{3}\mathbf{v}$ in Eq.(52) turns out to be proportional to $%
(1/v^{0})=\left( 1+\left\vert \mathbf{v}\right\vert ^{2}\right) ^{-1/2}$,
with no other dependence on the spatial component $\mathbf{v}$ of
four-velocity (See Eqs.(54) and (62)-(67)).

Next, denoting the three respective contributions to Eq.(52) of the three
terms of Eq.(36) by $\left\langle 0\right\vert \Delta H_{1}\left\vert
0\right\rangle $, $\left\langle 0\right\vert \Delta H_{2}\left\vert
0\right\rangle $, and $\left\langle 0\right\vert \Delta H_{3}\left\vert
0\right\rangle $, respectively, then%
\begin{equation}
\left\langle 0\right\vert \Delta H\left\vert 0\right\rangle =\left\langle
0\right\vert \Delta H_{1}\left\vert 0\right\rangle +\left\langle
0\right\vert \Delta H_{2}\left\vert 0\right\rangle +\left\langle
0\right\vert \Delta H_{3}\left\vert 0\right\rangle .
\end{equation}%
Then using Eqs.(36), (52), and (56), one has%
\begin{eqnarray}
&&\left\langle 0\right\vert \Delta H_{1}\left\vert 0\right\rangle  \\
&=&\frac{V}{32\pi ^{3}\hbar m}\rho _{0}^{4}\Delta ^{3}\mathbf{v}\left\langle
0\right\vert \int dv^{0}\delta \mathbf{(}\rho _{0}^{2}v^{2}-\rho
_{0}^{2})d^{3}\mathbf{x}\frac{\partial }{\partial x^{0}}\phi \left( \mathbf{%
x,-}v\right) \frac{\partial }{\partial x^{0}}\phi \left( \mathbf{x,}v\right)
\left\vert 0\right\rangle ,  \notag
\end{eqnarray}%
and substituting Eqs.(25) and (55) in Eq.(57), one obtains%
\begin{eqnarray}
&&\left\langle 0\right\vert \Delta H_{1}\left\vert 0\right\rangle   \notag \\
&=&\frac{V}{32\pi ^{3}\hbar m}\rho _{0}^{2}\Delta ^{3}\mathbf{v}\int \frac{%
dv^{0}}{2\left[ 1+\left\vert \mathbf{v}\right\vert ^{2}\right] ^{1/2}}\left[
\delta \left( v^{0}-\left[ 1+\left\vert \mathbf{v}\right\vert ^{2}\right]
^{1/2}\right) \right.   \notag \\
&&\left. +\delta \left( v^{0}+\left[ 1+\left\vert \mathbf{v}\right\vert ^{2}%
\right] ^{1/2}\right) \right] d^{3}\mathbf{x}\frac{d^{3}\mathbf{p}d^{3}%
\mathbf{p}^{\prime }}{\left( 2\pi \hbar \right) ^{3}2N\left( p^{0}p^{0\prime
}\right) ^{1/2}} \\
&&\left\langle 0\right\vert \left\{ -i\frac{p^{0}}{\hbar }e^{-ipx/\hbar
}e^{\rho _{0}pv/\hbar }\theta _{1}(-v^{0})a(\mathbf{p})+i\frac{p^{0}}{\hbar }%
e^{ipx/\hbar }e^{-\rho _{0}pv/\hbar }\theta _{1}(v^{0})a^{\dagger }(\mathbf{p%
})\right\}   \notag \\
&&\left\{ -i\frac{p^{0\prime }}{\hbar }e^{-ip^{\prime }x/\hbar }e^{-\rho
_{0}p^{\prime }v/\hbar }\theta _{1}(v^{0})a(\mathbf{p}^{\prime })+i\frac{%
p^{0\prime }}{\hbar }e^{ip^{\prime }x/\hbar }e^{\rho _{0}p^{\prime }v/\hbar
}\theta _{1}(-v^{0})a^{\dagger }(\mathbf{p}^{\prime })\right\} \left\vert
0\right\rangle .  \notag
\end{eqnarray}%
Here $\theta _{1}(\pm \rho _{0}pv/\hbar )$\ has been replaced by $\theta
_{1}(\pm v^{0})$, since it can be shown that $pv>mc$ when $v^{0}=\gamma >0$,
and $pv<mc$ when $v^{0}=-\gamma <0$ \cite{Brandt18}, \cite{Brandt19}. Then,
according to Eqs.(27), (30) and (54), one has%
\begin{equation}
\theta _{1}(v^{0})\theta _{1}(-v^{0})=0,\ \ \ \ [\theta
_{1}(v^{0})]^{2}=\theta _{1}(v^{0}).
\end{equation}%
Next using Eq.(59) in Eq.(58), one obtains%
\begin{eqnarray}
&&\left\langle 0\right\vert \Delta H_{1}\left\vert 0\right\rangle   \notag \\
&=&\frac{V}{32\pi ^{3}\hbar m}\rho _{0}^{2}\Delta ^{3}\mathbf{v}\int \frac{%
dv^{0}}{2\left[ 1+\left\vert \mathbf{v}\right\vert ^{2}\right] ^{1/2}}\left[
\delta \left( v^{0}-\left[ 1+\left\vert \mathbf{v}\right\vert ^{2}\right]
^{1/2}\right) \right.   \notag \\
&&\left. +\delta \left( v^{0}+\left[ 1+\left\vert \mathbf{v}\right\vert ^{2}%
\right] ^{1/2}\right) \right] d^{3}\mathbf{x}\frac{d^{3}\mathbf{p}d^{3}%
\mathbf{p}^{\prime }}{\left( 2\pi \hbar \right) ^{3}2N\left( p^{0}p^{0\prime
}\right) ^{1/2}}  \notag \\
&&\left\{ \frac{p^{0}p^{0\prime }}{\hbar ^{2}}e^{-i\left( p-p^{\prime
}\right) x/\hbar }e^{\rho _{0}\left( p+p^{\prime }\right) v/\hbar }\theta
_{1}(-v^{0})\left\langle 0\right\vert a(\mathbf{p})a^{\dagger }(\mathbf{p}%
^{\prime })\left\vert 0\right\rangle \right.   \notag \\
&&+\left. \frac{p^{0}p^{0\prime }}{\hbar ^{2}}e^{i\left( p-p^{\prime
}\right) x/\hbar }e^{-\rho _{0}\left( p+p^{\prime }\right) v/\hbar }\theta
_{1}(v^{0})\left\langle 0\right\vert a^{\dagger }(\mathbf{p})a(\mathbf{p}%
^{\prime })\left\vert 0\right\rangle \right\} .
\end{eqnarray}%
One also has%
\begin{equation}
a(\mathbf{p})\left\vert 0\right\rangle =0,\ \ \ \left\langle 0\right\vert
a^{\dagger }(\mathbf{p})=0,\ \ \left\vert \mathbf{p}\right\rangle
=a^{\dagger }(\mathbf{p})\left\vert 0\right\rangle ,\ \ \left\langle \mathbf{%
p|p}^{\prime }\right\rangle =\delta ^{3}(\ \mathbf{p}-\mathbf{p}^{\prime }).
\end{equation}%
Therefore using Eqs.(61) in Eq.(60), one obtains%
\begin{eqnarray}
&&\left\langle 0\right\vert \Delta H_{1}\left\vert 0\right\rangle   \notag \\
&=&\frac{V}{32\pi ^{3}\hbar m}\rho _{0}^{2}\Delta ^{3}\mathbf{v}\int \frac{%
dv^{0}}{2\left[ 1+\left\vert \mathbf{v}\right\vert ^{2}\right] ^{1/2}}\left[
\delta \left( v^{0}-\left[ 1+\left\vert \mathbf{v}\right\vert ^{2}\right]
^{1/2}\right) \right.   \notag \\
&&\left. +\delta \left( v^{0}+\left[ 1+\left\vert \mathbf{v}\right\vert ^{2}%
\right] ^{1/2}\right) \right] d^{3}\mathbf{x}\frac{d^{3}\mathbf{p}d^{3}%
\mathbf{p}^{\prime }}{\left( 2\pi \hbar \right) ^{3}2N\left( p^{0}p^{0\prime
}\right) ^{1/2}}  \notag \\
&&\frac{p^{0}p^{0\prime }}{\hbar ^{2}}e^{-i\left( p-p^{\prime }\right)
x/\hbar }e^{\rho _{0}\left( p+p^{\prime }\right) v/\hbar }\theta
_{1}(-v^{0})\delta ^{3}\left( \mathbf{p}-\mathbf{p}^{\prime }\right) ,
\end{eqnarray}%
or equivalently, 
\begin{equation}
\left\langle 0\right\vert \Delta H_{1}\left\vert 0\right\rangle =\frac{V^{2}%
}{32\pi ^{3}\hbar ^{3}m}\frac{1}{\left( 2\pi \hbar \right) ^{3}4N}\frac{\rho
_{0}^{2}\Delta ^{3}\mathbf{v}}{v^{0}}\int \frac{d^{3}\mathbf{p}}{p^{0}}%
p^{02}e^{-2\rho _{0}pv/\hbar }.
\end{equation}

Analogously, corresponding to the contributions to Eq.(56) of the two
remaining terms of Eq.(38), containing derivatives with respect to space and
four-velocity, respectively, one obtains%
\begin{equation}
\left\langle 0\right\vert \Delta H_{2}\left\vert 0\right\rangle =\frac{V^{2}%
}{32\pi ^{3}\hbar ^{3}m}\frac{1}{\left( 2\pi \hbar \right) ^{3}4N}\frac{\rho
_{0}^{2}\Delta ^{3}\mathbf{v}}{v^{0}}\int \frac{d^{3}\mathbf{p}}{p^{0}}%
\left\vert \mathbf{p}\right\vert ^{2}e^{-2\rho _{0}pv/\hbar },
\end{equation}%
and%
\begin{equation}
\left\langle 0\right\vert \Delta H_{3}\left\vert 0\right\rangle =\frac{V^{2}%
}{32\pi ^{3}\hbar ^{3}m}\frac{1}{\left( 2\pi \hbar \right) ^{3}4N}\frac{\rho
_{0}^{2}\Delta ^{3}\mathbf{v}}{v^{0}}\int \frac{d^{3}\mathbf{p}}{p^{0}}%
m^{2}c^{2}e^{-2\rho _{0}pv/\hbar }.
\end{equation}%
Then substituting Eqs.(63)-(65) in Eq.(56), noting that%
\begin{equation}
p^{02}=\left\vert \mathbf{p}\right\vert ^{2}+m^{2}c^{2},
\end{equation}%
evaluating the following integral \cite{Brandt24}, \cite{Ober},%
\begin{equation}
\int d^{3}\mathbf{p\ }p^{0}e^{-2\rho _{0}pv/\hbar }=\frac{\pi \hbar
^{2}m^{2}c^{2}}{\rho _{0}^{2}}K_{2}\left( 2\rho _{0}mc/\hbar \right) ,
\end{equation}%
where $K_{2}\left( z\right) $ is the modified Bessel function of the third
kind of order 2, substituting Eq.(28), and dividing by the spatial volume $V$%
, one finally obtains the vacuum energy density seen by an observer at $%
(x,v) $ in the spacetime tangent bundle:%
\begin{equation}
\frac{\left\langle 0\right\vert \Delta H\left\vert 0\right\rangle }{V}=\frac{%
m^{3}c^{4}}{32\pi ^{3}\hbar ^{2}\rho _{0}}\frac{K_{2}\left( 2\rho
_{0}mc/\hbar \right) }{K_{1}\left( 2\rho _{0}mc/\hbar \right) }\frac{\Delta
^{3}\mathbf{v}}{v^{0}}.
\end{equation}%
Thus any one observer sees a finite vacuum energy density given by Eq.(68).
One notes that for sufficiently small spread in the spatial part of the
observer's four-velocity (Eq.(53)), the observed vacuum energy density may
be near vanishing. This may be consistent with a near vanishing cosmological
constant.

\end{document}